\documentclass[12pt,a4paper]{article}
\usepackage{jheppub}
\usepackage{graphicx}
\usepackage{amssymb}
\usepackage{epsf,epsfig}
\usepackage{subcaption}

\abstract{In this chapter of the High Energy Physics Software Foundation Community Whitepaper, we discuss the current state of infrastructure, best practices, and ongoing developments in the area of data and software preservation in high energy physics. A re-framing of the motivation for preservation to enable re-use is presented.  A series of research and development goals in software and other cyberinfrastructure that will aid in the enabling of reuse of particle physics analyses and production software are presented and discussed.}

\title{HEP Software Foundation Community White Paper Working Group -- Data and
Software Preservation to Enable Reuse}

\author[a,1]{M. D. Hildreth,}
\author[b]{A. Boehnlein,}
\author[c]{K. Cranmer,} 
\author[d]{S. Dallmeier-Tiessen,}
\author[e]{R. Gardner, }
\author[f]{T. Hacker,}
\author[c]{L. Heinrich,}
\author[g]{I. Jimenez,}
\author[h]{M. Kane,}
\author[i]{D. S. Katz,}
\author[j]{T. Malik,}
\author[g]{C. Maltzahn,}
\author[k]{M. Neubauer,}
\author[l]{S. Neubert,}
\author[m]{Jim Pivarski}
\author[n]{E. Sexton-Kennedy,}
\author[d]{J. Shiers,}
\author[b]{T. Simko,}
\author[o]{S. Smith,}
\author[p]{D. South, }
\author[q]{A. Verbytskyi,}
\author[r]{G. Watts,}
\author[e]{J. Wozniak}

\affiliation[1]{Working Group Convener}
\affiliation[a]{University of Notre Dame, Notre Dame, IN, USA}
\affiliation[b]{Thomas Jefferson National Accelerator Facility, Newport News, VA, USA}
\affiliation[c]{New York University, New York, NY, USA}
\affiliation[d]{CERN, Geneva, Swizterland}
\affiliation[e]{University of Chicago, Chicago, IL, USA}
\affiliation[f]{Purdue University, West Lafayette, IN, USA}
\affiliation[g]{University of California Santa Cruz, Santa Cruz, CA, USA}
\affiliation[h]{Soundcloud, Berlin, Germany}
\affiliation[i]{NCSA, Urbana-Champaign, IL, USA}
\affiliation[j]{DePaul University, Chigago, IL, USA}
\affiliation[k]{University of Illinois Urbana-Champaign, Urbana-Champaign, IL, USA}
\affiliation[l]{University of Heidelberg, Heidelberg, Germany}
\affiliation[m]{Princeton University, Princeton NJ, USA}
\affiliation[n]{Fermi National Accelerator Laboratory, Batavia, IL, USA}
\affiliation[o]{McMaster University, Hamilton, Ontario, Canada}
\affiliation[p]{DESY, Hamburg, Germany}
\affiliation[q]{Max-Planck-Institut f{\"u}r Physik, Munich, Germany }
\affiliation[r]{University of Washington, Seattle, WA, USA}

\begin{document}

\maketitle

\newpage

\hypertarget{overview}{%
\section{Overview}\label{overview}}
As part of the process to produce a Community White Paper (CWP)\cite{CWP} within the efforts of the High Energy Physics Software Foundation (HSF)\cite{HSF}, a working group on Data and Software preservation was established to contribute to the broader discussion of how efforts in this area tied in with the overall goals of the HSF and the HEP community.  This document is essentially the chapter contributed to the CWP by this working group.

The Data and Software Preservation Working Group examined current preservation practices and technologies in HEP and beyond with the goal of describing a set of tools and procedures that will allow analysis preservation, discovery, and reuse. In addition, the group examined the issues around preserving the capabilities to read, simulate, and process the large experimental dataset from the raw data through physics analysis level data.  Clearly, the technical, sociological, and financial issues surrounding this topic are complex. This overview outlines the main points of the discussion, summarizes some interesting new developments, and suggests several areas of research and development that have the potential to advance the tools available for preservation as well as providing more of an impetus to do so.

Several recent documents produced by current projects in this area are readily available.  Because of this, a larger ``White Paper" on knowledge preservation seems redundant.  For more details, including use cases and a detailed description of the efforts within and among the various HEP experiments, the reader is urged to consult the 2015 DPHEP report\cite{DPHEP16}, documentation describing the CERN Open Data Portal\cite{CERNOpen} and the CERN Analysis Preservation Portal\cite{CAP}, the work of the DASPOS project\cite{DASPOS}, and a recent report on public access to data in the Mathematical and Physical Sciences in the United States\cite{MPSOpen}, which provides many other references.

\hypertarget{introduction}{%
\section{Introduction and Motivation}\label{introduction}}
It is easy to state a simple motivation for the preservation of the data and knowledge surrounding High Energy Physics (HEP) experiments. Given the large monetary investments that have been and will be made in particle physics experiments, it is incumbent upon physicists to preserve the data and the knowledge that can lead to scientific results in a manner such that this investment is not lost to future generations of analysts.  The word``knowledge" in this context includes processing and analysis software, documentation, and other components necessary for reusing a given dataset. This includes preserving processed data containing an appropriate level of detail to enable new analysis (``Level 3" data as defined in the 2009 DPHEP report\cite{DPHEP09} ). Preservation of this type can enable, for example, new analyses on older data, as well as a way to revisit the details of a result after publication.  Among other applications, the latter use can be especially important in resolving conflicts between published results, applying new theoretical assumptions, or evaluating different theoretical models.  Preservation enabling reuse offers specific benefits within a given experiment as well.  The preservation of software and workflows such that they can be shared enhances collaborative work between analysts and analysis groups, provides a way of capturing the knowledge behind a given analysis during the review process, enables easy transfer of knowledge to new students or analysis teams, and could establish a manner by which results can be generated automatically for submission to central repositories such as HEPData\cite{HEPDATA}, etc.  At a base level, preservation within an experiment could provide ways of re-processing and re-analyzing data that could have been collected more than a decade earlier. A final series of motivations comes from the potential re-use by others outside of the HEP experimental community.  Significant outreach efforts bringing the excitement of analysis and discovery to younger students has been enabled by the preservation of experimental data and software in an accessible format.  Many examples also exist of phenomenology papers reinterpreting the results of a particular analysis in a new context.  This has been extended further with published results based on the re-analysis of processed data by scientists outside of the collaborations\cite{MITQCD}.  Engagement of external communities, such as machine learning specialists, can be enhanced by providing the capability to process and understand low-level HEP data in portable and relatively platform-independent packages. This allows external users direct access to the same tools and data as the experimentalists working in the collaborations.  Connections with industrial partners, such as those fostered by CERN OpenLab\cite{Openlab} can be facilitated in a similar manner.  Note that the focus of the efforts described here is to both enable and facilitate knowledge preservation so that becomes easy to do.  Policy considerations on what workflows, analyses, and other knowledge should be preserved will necessarily be left to the scientists and the experiments to decide.

\hypertarget{WGOverview}{%
\section{Overview of Working Group Considerations}\label{WGOverview}}
In the process of evaluating what should be addressed within the scope of this working group, a consensus emerged on a set of elements that should form the basis of preservation systems that would enable HEP data to follow the FAIR principles as defined by the Force11 working group\cite{F11}, namely that they be findable, accessible, interoperable, and reusable.  Note that reusability implies, to a large extent, the other attributes, which is why we propose the goal of reuse.  Here a ``preservation system" is taken to mean a set of software tools coupled with a searchable archive that enables reuse. To the extent that they need to be developed, the elements that might comprise such a system are reflected in the research road map discussed below.  Throughout, the assumption is that the bit-wise preservation of data and other knowledge products, while neither a simple nor a zero-cost endeavor, is a solved problem in that items stored on digital media will continue to be retrievable with an acceptably small error rate long into the future.  Clearly, a variety of preservation systems will probably be appropriate given the different preservation needs between large central experiment workflows and the work of an individual analyst.  The larger question is to what extent common low-level tools can be provided that address similar needs across a wide scale of preservation problems.  The sections below contain some suggestions and examples of the kinds of elements that may be useful.

\subsection{Assessment Tools}\label{assessment}
When approaching the complex question of preservation, the question of ``what should be preserved to enable reuse?" must be answered\cite{MPS2}.  Of course, the answer depends on the level of reuse that is desired and the potential preservation framework in which the information will reside.  For any set of preservation systems, a set of guidelines should be developed that reflect the base assumptions on which preservation choices are made, that outline the features of each system, and that enable a user to choose a path toward preservation that is appropriate for the use case.  These guidelines or tools incorporating them can be complemented by different components of the preservation systems. One example could be a software tool executed at run time that is able to automatically provide the user with information on which external components the preserved elements depend. For example, to reuse a piece of analysis software, a potential new user may need access to data catalogues, conditions databases, software configurations, and maybe even experiment-specific software build systems. A tool that is able to observe the execution of the code before it is preserved could supply a list of these external elements, potentially while packaging the code up for preservation. A second example of a complementary tool is the design of elements that determine capture fidelity directly into the preservation framework.  These elements can audit captured information, data, executables, etc., to ensure that the stored elements are sufficient to satisfy the user's desired level of preservation.

\subsection{Capture Tools}\label{capture}
While a host of diverse tools exist for various aspects of digital preservation\cite{WflwTools}, very few provide a complete solution that enables scientists to package up their data, workflows, codes, executables, and documentation for preservation. And, none of these is suitable for the large-scale computational workflows typical of HEP.  The diversity of workflows, analysis practices, scales of computation, and computing techniques makes this a challenging area in which to suggest research.  It is likely that different tools may need to be developed to address the wide scale of preservation needs and scales within HEP and beyond.  It may be possible, however, to develop scalable infrastructure building-blocks that can be used to construct these tools.  No matter what forms these tools take, it is desirable that the capture of information, metadata, and the other necessary elements enabling preservation be done automatically, with as little effort from the user as possible.  

The simplest need that should be served by these tools is the ability to run a given executable,``wrap it up", and then re-run it at any future time while obtaining the same results.  The manner in which this is done and the execution environments could be quite varied.  Techniques such as continuous integration could be brought into analysis workflows, for example.  Changes to code, workflows, and data products would be tracked and tested automatically, leaving a complete record of their evolution.  This would naturally serve as a "continuous preservation" environment, assuming there are sufficient personnel to intervene and resolve any problems raised by the integration framework.   At the opposite end of the spectrum, yet still served by continuous integration, is the infrastructure necessary for full-scale production of processed events.  Since reuse of data often implies a comparison of new theoretical models, the infrastructure required to regenerate simulated events must also be preserved in order to insure that these comparisons are possible.  Currently, only minimal central efforts are underway to preserve the capabilities for data reprocessing or the full simulation chain in large HEP experiments. Looking 10 or even 5 years in the future, this may render the current data unusable.  The working group considered this a serious problem that should be addressed before the expertise to do so becomes sparse.  

While not directly related to capture tools, the availability and sustainability of infrastructure like CVMFS, which provides so many services to HEP experiments, is also a serious issue.  Can we be sure that CVMFS will provide the appropriate libraries needed to run older software versions several years from now?  Will CVMFS itself have a well-delineated future as computing platforms become more heterogeneous?  These are questions that should be addressed.

\begin{figure}[bthp]
    \centering
 \includegraphics[width=0.9\textwidth]{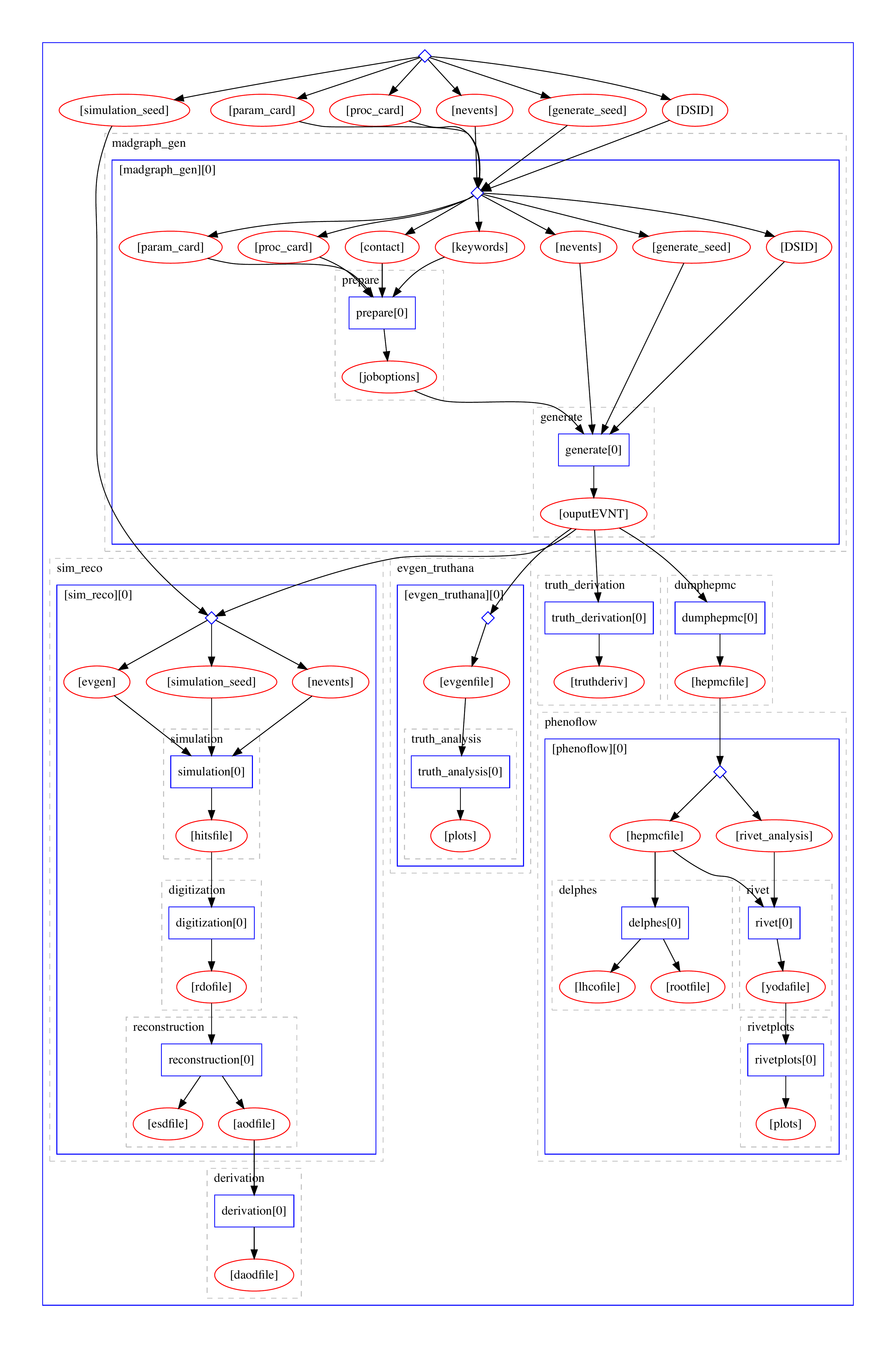}
    \caption{A sample REANA workflow, constructed from elements specified in Yadage, showing the different components and the order of operation.}
    \label{fig:workflow}
\end{figure}

\subsection{Applications: Preservation-as-a-Service}\label{Applications_PaaS}
\begin{figure}[bthp]
    \centering
 \includegraphics[width=0.9\textwidth]{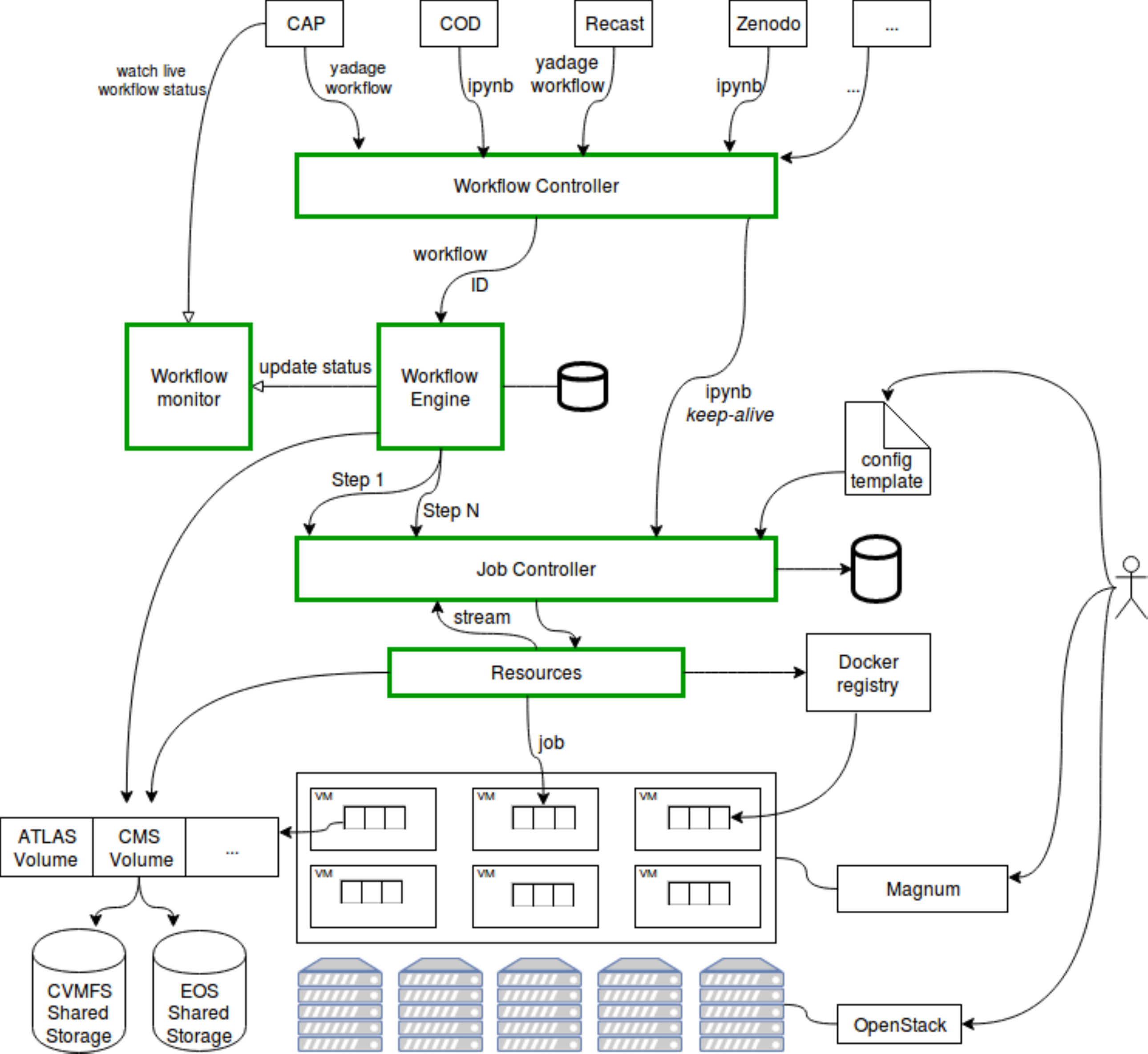}
    \caption{The REANA framework being developed as part of the CERN Analysis Preservation Portal.}
    \label{fig:REANA}
\end{figure}

In the space of knowledge preservation, several preservation systems have been or are being built that provide infrastructure that users can employ in order to share and reuse elements of workflows, data, etc. One of the first of these systems in HEP is RECAST\cite{RECAST}, which provides the capability to reinterpret the constraints an analysis may place on new physics by comparing new models against a what was found in a completed search analysis on experimental data.  The novelty here is that the RECAST infrastructure maintains the capability to re-run the full simulation and analysis chain given the new physics model as input. If run as a service for theorists, for example, the individual components of the analysis workflow are not accessible.  The re-analysis is performed as a``black box" process, and the results are subsequently made available. However, the technology for preserving, composing, and orchestrating the processing of the different workflow steps within RECAST has evolved to a more general system called REANA  that is under construction within/behind the CERN Analysis Preservation Portal (CAP)\cite{CAP}.  In this framework, an analyst can preserve individual workflow steps in containers and string them together with a series of scripts that enable quite complex workflows.  An example workflow represented by these scripts can be seen in Figure \ref{fig:workflow}.  Within REANA, the preserved executables can be reinstantiated according to the instructions in the workflow scripts by a job scheduler and workflow system that can be attached to any generic computing resource that supports container execution. Other means of instantiating executables, such as the more general approach represented by the Umbrella project\cite{Umbrella}, have also been added to this infrastructure.  A diagram of the REANA system is shown in Figure \ref{fig:REANA}. If appropriate, the executables representing individual workflow steps can be shared, swapped out for different processes, and re-used in different workflows.  The infrastructure also provides a container registry and will eventually have discovery tools that will enable users to find executables that are suitable for their purposes.  Finally, the infrastructure under construction is built entirely from commodity elements, making it possible to install at any site with appropriate computing capability.

A second type of preservation-as-a-service infrastructure is represented by common analysis and workflow execution frameworks such as the ALICE``train model". In these projects, individual users can include their analysis workflows in a central process that automatically runs their code as it makes a pass through the experimental dataset. This requires more effort up front, as the analysis codes need to be validated and placed in official repositories before the analysis train is run.  It does, however, provide preservation and versioning of the analysis code and a means of automatic execution, something that is unusual for analysis codes.

A third example of a preservation infrastructure is represented by the work done at DESY\cite{DESY} in building a system that automatically performed periodic validation of a given experimental code base that has been targeted for preservation.  As computing technology, software libraries, compilers, etc., are updated, standard validation scripts are run to assess if changes to the computing environment have caused changes in the physics results or outputs, or if work needs to be done to allow the code to even run after the underlying changes.  All major experiments now use such systems to track changes between software releases during development.  Almost none use these techniques to monitor the viability of older code and software releases. 

\hypertarget{Policies}{%
\section{Current Policies of the LHC Experiments}\label{Policies}}

Each of the LHC experiments has adopted a data access and/or data preservation policy, all of which can be found here.  All of the LHC experiments support public access to some subset of the data in a highly-reduced data format for the purposes of outreach and education.  CMS has gone one step further, releasing substantial datasets in an AOD format that can be used for new analysis.  The data release includes simulated data, virtual machines that can instantiate the added analysis examples, and extensive documentation\cite{CMSOpenData}.  ALICE has promised to release 10\% of their processed data after a five-year embargo and has released 2010 data at this time\cite{ALICEOpenData}.  LHCb has promised to release 50\% of the processed data and associated software after five years contingent on having sufficient manpower to produce this.  So far, that has not been the case. ATLAS has chosen a different direction for data release, as is evidenced by this excerpt from their public access policy\cite{ATLASOpenData}: 

\begin{quote} ``Data associated with journal publications are also made available: tables and data from plots (e.g. cross section values, likelihood profiles, selection efficiencies, cross section limits, ...) are stored in appropriate repositories such as HEPDATA\cite{HEPDATA}. ATLAS also strives to make additional material related to the paper available that allows a reinterpretation of the data in the context of new theoretical models. For example, an extended encapsulation of the analysis is often provided for measurements in the framework of RIVET\cite{Rivet}. For searches information on signal acceptances is also made available to allow reinterpretation of these searches in the context of models developed by theorists after the publication. ATLAS is also exploring how to provide the capability for reinterpretation of searches in the future via a service such as RECAST. RECAST allows theorists to evaluate the sensitivity of a published analysis to a new model they have developed by submitting their model to ATLAS."
\end{quote}

None of the LHC experiments have made recent public statements addressing the new capabilities of the CERN Analysis Portal and whether or not some use of it will be required (or strongly encouraged).  All of them support some mechanisms for internal preservation of the knowledge surrounding a physics publication.

\hypertarget{Challenges}{%
\section{Key Challenges for Preservation and Reuse}\label{challenges}}
Creating the infrastructure and tools that will allow HEP data to adhere to the FAIR principles of being findable, accessible, interoperable, and, reusable presents several challenges, which are outlined in the sections below.
\subsection{Interoperability}\label{Interoperability}
Almost all of the systems and techniques discussed in the previous section rely on the capture of executables and their preservation for reuse. Today, the optimal technology to facilitate this is the linux container.  Even the systems such as the analysis train model which start from preserved code are generally compiled for a small number of hardware platforms.  However, with the proliferation of computing architectures and the potential customization required for optimal performance, it may be quite difficult to guarantee that any preserved executable can be run on an arbitrary compute node.  The large HEP experiments have already encountered many difficulties in this area when trying to use large HPC facilities in an opportunistic fashion.  As the computing hardware becomes more heterogeneous, new solutions may need to be found for sustainable computing.

An alternate approach would be to ensure that whatever tools are developed for executable and workflow capture function on most platforms where computing is done.  This would guarantee that preservation could be done at the time of execution. Of course, all such systems will face the challenges of tracking the evolution of the underlying hardware once they function at one point in time.

No matter what preservation tools are developed that might enable reuse of software, analysis techniques, and data, if they are not conceived from the beginning as an integral part of the standard frameworks, retrofitting will be nearly impossible.  Much of the information that is needed to enable reuse is transient and available only at the time of processing.  For example, an analysis framework with embedded preservation infrastructure that enables the automatic capture of appropriate details and metadata describing the workflows, the data processed, and the output products would allow the user to compile a record of that computation and preserve it for later reuse, archiving, or sharing with others.  Without automatic capture, assembling the necessary information is painful and time-consuming, which dramatically raises the effort for preservation. 

\subsection{Findability}\label{Findability}
For reuse within an experiment or reuse more broadly, the elements that have been preserved need to be linked with some sort of metadata record in a searchable database. While simple search-able parameters can be attached to data and software objects, reuse places much more stringent requirements on documentation and demands a sophisticated metadata description for data objects, executables, workflow steps, and workflows themselves.  Substantial work has been done in this area in order to provide suitable metadata descriptions of computation, workflows, etc.\cite{Ontology} which will be incorporated into the CERN Analysis Preservation Portal, but the current situation represents a work in progress rather than a fully-developed solution.  Further investment in metadata schema implementation, prototyping, and testing will be required before a suitably-descriptive metadata vocabulary can exist.
 
Once a metadata description that is suitably complex to enable re-use is in place, a registry or database structure will also be required in order to both store analysis elements and to enable users to search for and find them.  While the effort to create such an infrastructure is probably small compared to that needed for the metadata description, it is still a necessary part of the system enabling preservation and reuse.

\subsection{Cost Justification}\label{Cost}
Clearly, adding the infrastructure necessary for preservation and reuse will require a substantial investment in development and, to a lesser extent, hardware for the implementation.  As discussed above, the motivations for doing so are strong.  The premise of this document is that the motivations do indeed justify the cost of development, especially since many of the basic elements of the entire preservation and reuse ecosystem already exist or are currently being envisioned, designed, and built. As the complexities of computational workflows increase and become increasingly ubiquitous across science, there is a exponentially growing need for the sorts of tools and infrastructure described here.
 
There is, however, a difference between mandated re-use and reproducibility and tools that enable these yet grow organically from the needs of the analysts. If an infrastructure that enables preservation and reuse can be embedded in the frameworks and analysis procedures in a way that makes the analyst's work more productive, then there is little cost to the analyst beyond the adoption of the analysis framework.  Increased productivity benefits everyone, resulting in more rapid production of results.  The barriers to preservation are lowered by making it automatic, and automatically useful to the individual physicist and the larger collaborations. Focusing on the internal benefits of such systems may be the simplest means to accelerate their acceptance.

\subsection{Adoption Challenges and Incentive Structure}\label{Adoption}
An additional challenge is the actual adoption of preservation tools and techniques by the community of physicists. An advantage to embedding them in a framework that is the implementation of a completely new analysis model is that they are there from the start.  As discussed elsewhere, many of the standard techniques of software, such as continuous integration, could have a profound impact on the way analysis is done as well as providing the means for automatic preservation.  Hence, efforts to design and implement new models of data analysis offer a unique opportunity to realize the benefits of preservation, sharing, and reuse as long as these elements are considered as integral parts of the infrastructure.  

There has been much discussion about what sorts of modifications might be necessary to the scientific and academic incentive/reward structure in order to create a ``culture of preservation."  While such cultural and institutional changes might drive the adoption of preservation techniques and make re-use of scientific software and data widespread, a top-down approach is difficult to implement and to ensure that it meets the needs of the community.  An effort originating in the community itself, however, and one that makes the community more productive may well be the path of least resistance.

\hypertarget{Connections}{%
\section{Connections with Other Working Groups}\label{Connections}}
Aspects of preservation and reuse underlie many of the themes discussed across a wide variety of working groups whose conclusions appear in the CWP report. Some major overlapping themes are discussed here.

\subsection{Data Analysis and Interpretation WG}
One focus of the Data Analysis and Interpretation Working Group[{\bf cite report}] is designing an analysis framework that enables the analyst to remember and reproduce a piece of an analysis or an entire workflow at some time removed from its original creation. Here the emphasis is not so much on scientific integrity but just on being able to pause and reliably restart an analysis effort. Creating an infrastructure with this capability would incorporate essentially all aspects of a larger preservation system including: provenance tracking of results (which jobs ran when with which input and produced which output?), a means of capturing an executable and its computation environment, and some method of cataloguing the results.  A system of this type may be sufficiently modular to allow the arbitrary composability of different analysis executables, as seen in the REANA framework.  Clearly, the proposed elements of this framework would be very much in alignment with the broad outline of infrastructure discussed in this document. Many of the building blocks that need to be developed can be both widely applicable and shared among preservation systems.

\subsection{Other Groups}
Other elements of preservation and reuse infrastructure are clearly relevant for other working groups within the CWP effort.  For example, tools for executable and environment packaging, workflow preservation and reinstantiation have obvious overlaps with the interests of the Workflow and Resource Management 
WG\footnote{For the detailed overview and statements of each of the WGs, please see Reference \cite{CWP} and the individual WG chapters therein.} and the Facilities and Distributed Computing WG.  To the extent that continuous integration and other frameworks can contribute to software preservation, there is a direct connection to the Software Development WG.  Engagement with the Event/Data Processing Frameworks WG will be required in order to address how the production software for the experiments can be included in preservation frameworks for long-term use.  The Data Access, Organization, and Management WG clearly overlaps with long-term preservation as well.

\hypertarget{Roadmap}{%
\section{Meeting the Challenges: R\&D Roadmap}\label{Roadmap}}
While much of the infrastructure discussed in this document can be built today from existing technologies, several topics for research and development emerged from the discussions.  These are outlined below, followed by a roadmap for the next five years that will allow the exploration of many of these issues.  In proposing the R\&D roadmap, we note that, for many of the issues discussed, closer collaboration with industrial partners is absolutely necessary.  Not only have those in industry thought about many of these problems, it is likely that some of them have already developed paths to solutions.  The rapid advancement and proliferation of various container technologies is just one example of many. Whether these particular solutions are appropriate for HEP and other scientific applications is an open question and will need to be resolved on a case-by-case basis.

One issue that has broad implications for many of the projects in this report is the projected longevity of executables stored in containers.  This is, of course, directly related to the rapid evolution of computing architectures.  A series of important questions in this domain are currently unanswered and could form the basis of an important line of research.  For example, how much can one ``future-proof" executables and software against hardware evolution?  Is there an invariant set of operating system ``hooks" (similar to the POSIX standard\cite{POSIX}) that are necessary to guarantee that the executable within a container will still run some years into the future no matter what the changes in the hardware base?  Will it be possible to``nest" the execution of containers in VMs (in VMs) that preserve execution environments on new platforms? If so, what are the compromises in performance?

A second, related issue is one of software design and implementation, and the question of software sustainability in the modern era of heterogeneous computing architectures.  At what level is it possible to abstract the functional needs of a piece of software from the highly-optimized implementation that is required for computational speed and optimum memory use on a given hardware platform?  Is it possible (or desirable) to design low level libraries that can offer these implementations in a modular fashion, rather than at the level of single instructions?  One possible solution would be``toolkits" of numerical function libraries (e.g. a 21st century CERNLib) that are optimized for a particular class of hardware and that could be called from higher level code. Compilation systems could automatically recognize the platform (or kernel version) and choose the appropriate optimizations. 

As mentioned above, an area of investigation that links preservation and reuse with questions raised by the Computing Models, Facilities, and Distributed Computing WG is one of container orchestration.  Clearly, it will be necessary in the future to distribute and execute multiple different containers for large multi-step HEP workflows.  While many container orchestration tools have been developed (e.g., Kubernetes, Helios), do they scale to the size of HEP production needs, what new infrastructure is required in order to run multi-step workflows within a typical HEP production system, and how can all of the elements of the system be preserved and catalogued for re-use?

The evolution of analysis techniques to include``big data" applications such as Apache Spark based on databases or dedicated big data "appliances" raises a host of new issues for preservation and reuse.  Analysis done in this environment follows a different paradigm from the skim/slim/ntuple processes that has dominated HEP analysis for decades.  However, there has been significant development of HEP-specific analysis tools on these platforms\cite{DIANA}.  In principle, analysis scripts or their sub-components can be reused, if they are preserved in a discoverable manner along with the interfaces and middleware that allow them to function. A separate question concerns the data itself.  In principle, one ingests or formats an entire dataset in order to prepare it for analysis.  Since there is no reduced data to associate with the final results, a snapshot of the full dataset may be the most appropriate to designate as the basis for the analysis.  This becomes extremely unwieldy, since either the entire (large) dataset in its "ingested" state needs to be kept, or the procedure by which it is ingested (and the hardware and database configuration on which it was analyzed) needs to be reproducible. Clearly, time-dependent datasets present additional challenges. The coupling of the data and the underlying storage hardware in this paradigm and its implications for preservation and reuse need to be better understood.

One concrete contribution that an R\&D effort in this arena can make is the development of tools that facilitate preservation and reuse.  The high-level goals addressed here require low-level tools that individuals can use to easily capture their executable and a description of their computational environment, for example, or their multi-step workflow. Currently, these common tools do not exist in the HEP domain. In order to make progress in this area, various use cases need to be developed so that tools can be tailored to a particular task. An example set of these have already been enumerated in the context of developing CAP/REANA\cite{CAPExamp}.  Possible uses include those mentioned in the introduction, such as having a complete set of code and documentation for internal review of an analysis, evaluating new theoretical models using the results of old analyses, preserving a benchmark analysis for posterity or to pass to a new student, etc.  Once several use cases are established, the base or common elements that need to be captured and described with metadata can be extracted and generic tools created.  Ideally, this would lead to metadata and provenance capture infrastructure in addition to the preservation of the computations themselves.  With diverse use cases, tools that are applicable to more than one discipline can be developed.  In the case of HEP, the CAP/REANA infrastructure under construction at CERN can be exploited as a test bed for different capture tools and discoverability, making it the ideal development platform.

A potential second instance of development to fill a needs gap is one that targets validation infrastructure.  As mentioned above, despite the work done at DESY directed at the continual evaluation of the viability of a preserved code base, none of the current HEP experiments have implemented such a system.  This may be more a question of motivation rather than technical challenge.  It is an important issue, however, if a given experiment wishes to preserve the capability to run the simulation software for an old detector configuration in order to reinterpret an existing result or to re-analyze ``old" data. An initial investment in what would amount to a small extension of the standard continuous integration and validation infrastructures that the experiments already possess could potentially result in a dramatic reduction in the effort required to resuscitate past software versions.  

This leads to the larger question of the preservation of the production workflows and executables of the HEP experiments.  As mentioned above, none of the LHC experiments have projects to preserve the ability to re-use old software releases.  This presents a major obstacle to the re-use of the data for physics analysis, since, without further action, it will be impossible to re-run the detector simulation to understand the response of the detector to new physics signals.  While the infrastructure to capture the executables and the standard workflows exists, the lifetime of the candidate technologies is unknown.  That places this question in line with the rest of the R\&D topics presented in the above paragraphs.  There is an added complexity, here, in the external conditions databases and other external resources that are used to provide constants to the simulation that must also be sustained or separately preserved. 

\subsection{Research and Development Roadmap}\label{RandD}
Here, we present a progression of research topics that address the issues discussed above.  Note that many of them should be orchestrated in conjunction with projects conducted by the R\&D programs of other working groups, since the questions addressed are common.  The timescales indicated extend from the start of the program.

\smallskip
\noindent{\bf Near term: (1 year)}
\begin{itemize}
\item 
Demonstrate a fully-functional version of the CERN Analysis Preservation Portal (CAP), the container-based solution for workflow and executable preservation.  This includes the creation of a UI that allows a user to reproduce a workflow with a single command, the implementation of a metadata schema that allows the stored workflows and executables to be discovered with a search engine, the implementation of this search engine, and the provisioning of a scalable back-end system for computation.  Demonstrate composing new workflows from preserved executable steps.
\item 
Demonstrate the capability to provision and execute production workflows for LHC experiments that are composed of multiple independent containers.  Investigate the modifications necessary to the underlying WLCG and workflow management infrastructure in order to construct and deploy these workflows. Begin work on large-scale executable/workflow preservation.
\item
Begin the investigation of limitations on the longevity and ubiquity of container-based solutions to executable preservation.
\item
Collection of analysis use cases and which elements are necessary to preserve to enable re-use.  Begin the creation of a taxonomy of information that should be captured during analysis steps.  Determine if there are analysis workflows that are so different that they represent examples that cannot be captured in a format amenable to CAP.
\item
Begin investigation of preservation issues in ``big data" analysis environments.
\end{itemize}

\noindent{\bf Medium Term: (3 years)}
\begin{itemize}
\item
Develop prototype analysis ecosystem(s), including embedded elements for the capture of preservation information and metadata, and tools for the archiving of this information.
\item
Evaluation of prototypes and suitability for re-use: determination of which elements are missing in order to enable re-use, which elements are difficult to capture.  Needs assessment in order to determine future R\&D necessary to evolve prototypes.
\item
Continue evaluation of the limits of container technologies in the preservation arena.
\item
Continue exploration of large-scale analytic workflows.
\item
Develop prototype for the preservation and validation of large-scale production executables/workflows.
\end{itemize}

\noindent{\bf Longer Term: (5 years)}
\begin{itemize}
\item
Deploy analysis ecosystem that enables re-use for any analysis that can be conducted in the ecosystem.
\item
Deploy system for the preservation and validation of large-scale production executables/workflows
\end{itemize}

\hypertarget{engage}{%
\section{Engaging Communities Beyond HEP}\label{engage}}
Clearly, the problems discussed in this document are common to many applications in other areas of science; many industrial applications may also present solutions or suggest ways forward to arrive at solutions. A particular example of this is represented by the series of tools and procedures developed by the DevOps community.  The modification and adoption of some of these and their application to physics analysis would naturally solve a series of problems related to versioning and workflow preservation.  Beyond this, many examples also exist of, for example, workflow systems in other scientific fields.  Many may not scale to LHC-sized applications, but the system architectures can be quite similar to those developed in HEP.  Unfortunately, few opportunities exist to discover and exploit commonalities in conceptualization and implementation.  This is one case where the strong HEP commitment to these technologies could serve as a basis for reaching out to other scientific disciplines to foster the creation of communities of development centered around the advancement of this work.


\newpage

\begin{thebibliography}{99}
\bibitem{CWP} 
  A.~A.~Alves, Jr {\it et al.},
  {\it ``A Roadmap for HEP Software and Computing R\&D for the 2020s,''}
  arXiv:1712.06982 [physics.comp-ph].
\bibitem{HSF} {\tt http://hepsoftwarefoundation.org/}
\bibitem{DPHEP16} S. Amera, {\it et al.}, {\it ``Status Report of the DPHEP Collaboration: A Global Effort for Sustainable Data Preservation in High Energy Physics,''} arXiv:1512.02019v2.
\bibitem{CERNOpen} Accessible at {\tt http://opendata.cern.ch/about} .
\bibitem{CAP} Accessible at {\tt https://analysispreservation.cern.ch/ } .
\bibitem{DASPOS} Information available at {\tt daspos.crc.nd.edu} .
\bibitem{MPSOpen} {\tt  doi:10.7274/R0542KN7}; information available at {\tt mpsopendata.crc.nd.edu} .
\bibitem{DPHEP09} Z. Akopov, {\it et al.}, {\it ``Status Report of the DPHEP Study Group: Towards a Global Effort for Sustainable Data Preservation in High Energy Physics''}, {\tt https://arxiv.org/ftp/arxiv/papers/0912/0912.0255.pdf} .
\bibitem{HEPDATA} {\tt http://hepdata.cedar.ac.uk/} .
\bibitem{MITQCD} J. Thaler et al., {\tt 10.1103/PhysRevD.96.074003} .
\bibitem{Openlab} {\tt http://openlab.cern} .
\bibitem{F11} M. D. Wilkinson, {\it et al.}, {\it ``The FAIR Guiding Principles for scientific data management and stewardship,''} Scientific Data {\bf 3}, 160018 (2016).  {\tt doi:10.1038/sdata.2016.18}.  Information available at {\tt https://www.force11.org/group/fairgroup/fairprinciples} .
\bibitem{MPS2} For a lengthy discussion, see, for example, https://mpsopendata.crc.nd.edu/index.php/final-report.
\bibitem{WflwTools} See, for example, the COPTR compilation at {\tt http://coptr.digipres.org/Main\_Page} or the POWRR Tool grid: {\tt http://digitalpowrr.niu.edu/tool-grid/} .
\bibitem{RECAST} Kyle Cranmer and Itay Yavin, {\it ``RECAST: Extending the Impact of Existing Analyses,''} JHEP 1104:038, 2011; {\tt https://arxiv.org/abs/1010.2506} .
\bibitem{REANA} {\tt http://reanahub.io/ }.
\bibitem{Umbrella} {\tt https://ccl.cse.nd.edu/software/umbrella/} .
\bibitem{DESY} Dmitri Ozerov and David M South, 2014 J. Phys.: Conf. Ser. {\bf 513} 042043.
\bibitem{CMSOpenData} {\tt http://opendata.cern.ch/research/CMS} .
\bibitem{ALICEOpenData} {\tt http://opendata.cern.ch/education/ALICE} .
\bibitem{ATLASOpenData} {\tt https://cds.cern.ch/record/2002139/files/ATL-CB-PUB-2015-001.pdf} .
\bibitem{Rivet}  {\tt http://rivet.hepforge.org/} .
\bibitem{Ontology} See, e.g., D. Carral {\it et al.}, "An Ontology Design Pattern for Particle Physics Analysis", Proceedings of the 6th Workshop on Ontology and Semantic Web Patterns (WOP 2015) co-located with the 14th International Semantic Web Conference (ISWC 2015), Bethlehem, Pensylvania, USA, October 11, 2015.
\bibitem{POSIX} {\tt http://standards.ieee.org/develop/wg/POSIX.html} .
\bibitem{DIANA} The DIANA-HEP project;  {\tt http://histogrammar.org/} .
\bibitem{CAPExamp} {\tt http://doi.org/10.5281/zenodo.33693} .


\end{thebibliography}
 \end{document}